\documentclass[]{spie}  
\usepackage[]{graphicx}
\usepackage{amsmath,graphics,epsfig,mathrsfs}
\usepackage{bm}
\usepackage{epigraph}
\newcommand{\etal}{{\em et al.~}}

\title{Manipulating the Voltage Dependence of Tunneling Spin Torques}

\author{A. Manchon
\skiplinehalf
Physical Science and Engineering, King Abdullah University of Science and Technology (KAUST), Thuwal 23955, Saudi Arabia}

\authorinfo{Further author information: E-mail: aurelien.manchon@kaust.edu.sa, Telephone: (+966)544700061, Website: http://spintronics.kaust.edu.sa}

\begin{document}
  \maketitle

\begin{abstract}
Voltage-driven spin transfer torques in magnetic tunnel junctions provide an outstanding tool to design advanced spin-based devices for memory and reprogrammable logic applications. The non-linear voltage dependence of the torque has a direct impact on current-driven magnetization dynamics and on devices performances. After a brief overview of the progress made to date in the theoretical description of the spin torque in tunnel junctions, I present different ways to alter and control the bias dependence of both components of the spin torque. Engineering the junction (barrier and electrodes) structural asymmetries or controlling the spin accumulation profile in the free layer offer promising tools to design efficient spin devices. 
\end{abstract}



\section{INTRODUCTION}
\label{sec:intro}  

While most of the commercial microelectronic devices are based on the charge of the carriers (electrons and holes), Spintronics relies on the spin angular momentum of the carrier to generate low energy consumption functional devices. The most impressive demonstration of the technological relevance of Spintronics is the implementation of Giant Magnetoresistive read heads in magnetic data storage \cite{reviewchappert}. As illustrated by the current state of the art of data storage, the first virtue of magnetic devices is their non-volatility: in the absence of external input and thermal activation, the magnetic state remains frozen in its equilibrium direction. Therefore, non-volatile Magnetic Random Access Memories (MRAM) composed of magnetic tunnel junctions are expected to present outstanding performances down to 30nm and below.\par

There are currently various types of MRAM. The original MRAM concept, commercially available, is based on field-induced switching but has limited performances when scaling down the device area \cite{ieee}. Another candidate is the thermally-assisted MRAM, or TAS-RAM, where field-induced switching is assisted by thermal activation \cite{tas}. Beyond these different alternatives, spin transfer-MRAMs (STT-MRAMs) present a particular interest for both applied and fundamental condensed matter physicists \cite{ieee} through the electrical manipulation of the magnetic configuration of the device. This has been made possible by the prediction of the so-called spin transfer torque (STT) by Slonczewski and Berger \cite{Slonc96}: A spin-polarized current impinging on a ferromagnet can transfer its spin degree of freedom to the local magnetization exerting a torque on it. This effect has been observed in metallic spin-valves (two ferromagnetic layers separated by a spacer), tunneling junctions and even magnetic domain walls and is shown to induce current-control magnetic excitations, GHz self-sustained magnetic precessions and magnetization switching\cite{review1,reviewmtj,chapter}. This area has known tremendous developments in the past ten years with the realization of efficient devices such as spin torque MRAMs, radio-frequency oscillators and reprogrammable logics. \par

Beyond technological challenges, spin transfer torque in magnetic tunnel junctions (MTJs) presents a number of puzzles. Although STT in MTJs is considered to have reached maturity, the bias dependence extracted from routine measurements is still not well understood nor controlled. In this article, I propose an overview of the means to control the bias dependence of STT in MTJs. I present first the nature of the spin torque expected in an "ideal" ballistic tunnel junction. Then, I briefly review the influence of asymmetries and interfacial scattering and finally I discuss the influence of spin diffusion in the free layer.

\section{STATE OF THE ART: EXPERIMENTS}

Spin transfer torque in tunnel junctions has been first demonstrated in MgO-based junctions by Sun and in AlOx-based junctions by Huai et al. and Fuchs et al. \cite{huai}. However, Fe/MgO/Fe-type junctions very rapidly supplanted the AlOx junctions due to their outstanding transport properties. Since then, a number of exciting realizations have been achieved including switching, excitations, precessions, spin diode etc. We refer the reader to Ref. [6] for more details about the experimental achievements. Among the most important results, it has been demonstrated experimentally \cite{sankey} and theoretically \cite{theo,slonc07,manchon,wil,xiao,heiliger,tang} that the spin torque has the following form
\begin{eqnarray}\label{eq:1}
&&{\bm T}=T_{||}{\bm m}\times({\bm p}\times{\bm m})+T_\bot{\bm m}\times{\bm p},\\
&&T_{||}=a_1V+a_2V^2,\; T_\bot=b_0+b_2V^2.\label{eq:2}
\end{eqnarray}
Here, ${\bm m}$ and ${\bf p}$ are the magnetization direction of the free and pinned layers, respectively and $V$ is the bias voltage applied across the junction. The first term $T_\|$ in Eq. (\ref{eq:1}), referred to as the in-plane torque, competes with the magnetic damping allowing for self-sustained magnetic precessions and switching, whereas the second term $T_\bot$, referred to as the out-of-plane torque, acts like an effective field applied along ${\bf p}$. As we discuss in the next section, this dependence is theoretically predicted in ballistic junction when the transverse spin density is absorbed at the interface.\par

However, recent experiments have reported important discrepancies between the actual bias dependence of the spin torque and the one proposed in Eqs. (\ref{eq:1})-(\ref{eq:2}) \cite{deac,petit,li,oh}. In particular, Oh {\em et al.} \cite{oh} showed that in an asymmetrically designed MTJ, the bias dependence of the out-of-plane torque acquires a linear contribution. This linear contribution has also been observed by Petit et al. and Heinonen et al. \cite{petit} as small biases. Deac et al.\cite{deac} and Li et al.\cite{li} have also uncovered complex bias dependences at large bias voltages. The aim of the present article is to present a coherent description of the spin transfer torque in tunnel junctions in order to account for these different observations.\par

\section{BALLISTIC INTERFACIAL SPIN TORQUE}
\subsection{Ballistic Tunneling}

\begin{figure}[h!]
\centering
\includegraphics[width=12cm]{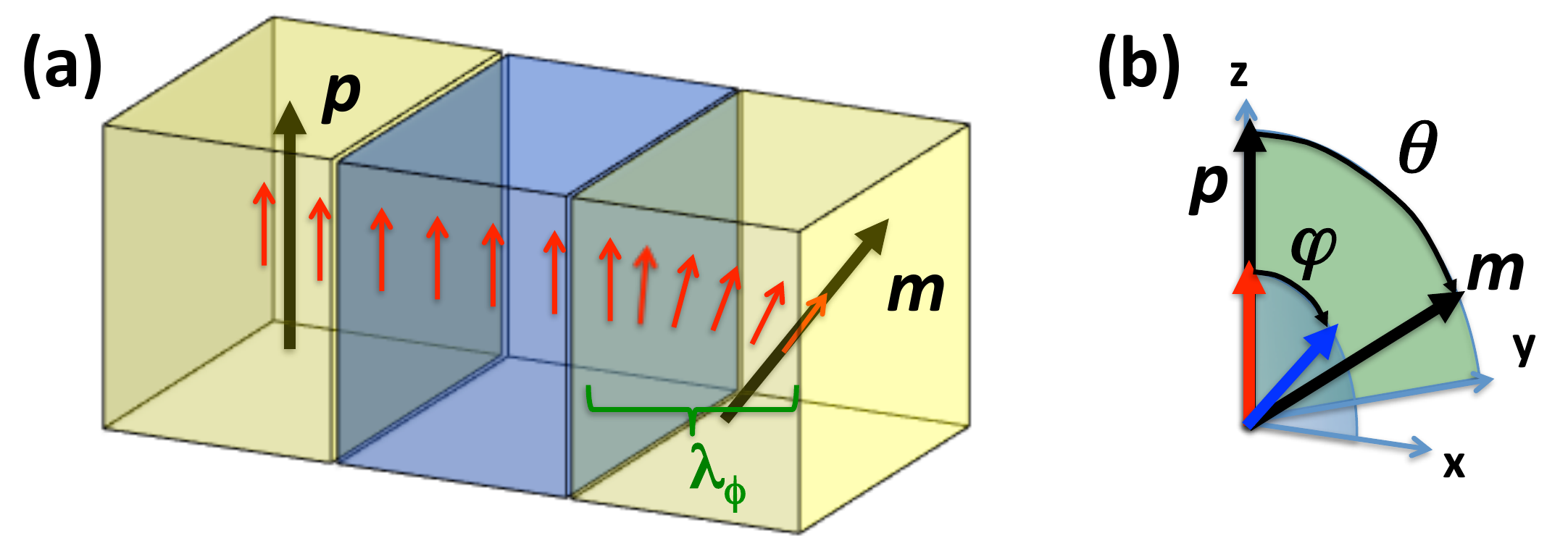}
\caption{\label{fig:fig1}(Color online) (a) Schematics of the spin transfer process in a magnetic tunnel junction. The black arrows refer to the magnetizations direction, and the red arrows represent the itinerant electron spins. Note that the itinerant spins get aligned on the local magnetization of the free layer ${\bm m}$ over a distance $\lambda_\phi$, called spin dephasing length; (b) Schematics of the alignment of the impinging spin in the {\em rigid spin} picture (red arrow) and in the realistic transport description (blue arrow). This supplementary angle gives rise to the out-of-plane torque.}
\end{figure}

At this early stage of the theory, we directly connect the spin torque acting on the volume of the free magnetic layer to the interfacial spin current transverse to the local magnetization, $T_{||(\bot)}={\cal J}_s^{x(y)}|_{interface}$. By definition, the spin transfer torque is due to the amount of spin angular momentum transferred to the magnetic layer, minus the amount of spin lost through spin-flip mechanisms. In the present case of ballistic transport (no spin relaxation), incoming spins gradually align on the local magnetization by transferring its transverse momentum to it, as schematically depicted in Fig.\ref{fig:fig1}(a). This alignment takes place on a very small distance near the interface, named the spin dephasing length $\lambda_\phi$. Assuming that the spin torque is related to the {\em interfacial} spin current means that the thickness of the free layer is much larger than the spin dephasing length ($d>>\lambda_\phi$). As we will discuss in the following, this definition is not appropriate in the general case and quantum resonances as well as spin diffusion effects can significantly affect the torque in the case of ultra thin free layers. Nonetheless,  we assume for now that the spin transfer is interfacial and given by the interfacial tunneling spin current.\par

The available theoretical descriptions of the bias dependence of spin torque in tunnel junctions have been obtained through transfer matrix formalism \cite{slonc07}, free electron \cite{xiao,wil,manchon,prb}, tight binding \cite{theo,tang} and ab-initio calculations \cite{heiliger}. Although these methods capture some or most of the band structure details of the system, they do not provide conceptual pictures about the symmetries of the bias dependence of the spin torque. In the present paragraph, we wish to provide such a qualitative description, disregarding any band structure details. A convenient method is the Bardeen Transfer Matrix (BTM) approach \cite{bardeen}, that was widely used by Slonczewski in his descriptions of spin torque in tunnel junctions \cite{slonc07}. The tunneling transport is expressed as the product of the interfacial densities of states through a transfer matrix. The simplest version of this model proposed by Julliere \cite{julliere} had a reasonable success in explaining the tunneling magnetoresistance. Formally, BTM approach can actually be recovered as an asymptotic limit of the Keldysh tight-binding model in the case of thick and large barriers \cite{nozieres}: As long as the interface states are {\em decoupled} (tunneling is treated perturbatively), BTM is applicable. Therefore, the present discussion does not reproduce quantitatively the physics of ultra thin barrier MTJs, but we believe it does provide the relevant symmetries and tendencies. In the spin-dependent BTM formalism, the charge and spin current densities are given in the spinor form in the $2\times2$ spin space
\begin{eqnarray}\label{eq:5}
\hat{J}=2\pi\frac{e}{\hbar}\int d\epsilon[\hat{\rho}_{L}\hat{T}_{L\rightarrow R}\hat{\rho}_{R}\hat{T}_{L\rightarrow R}^\dagger f_L(1-f_R)-\hat{\rho}_{R}\hat{T}_{R\rightarrow L}\hat{\rho}_{L}\hat{T}_{R\rightarrow L}^\dagger f_R(1-f_L)],
\end{eqnarray}
where $f_{L,R}$ and ${\hat \rho}_{L,R}$ are the Fermi distribution function and electronic density of states at the left (right) interfaces, and ${\hat T}_{L\rightarrow R}$ (${\hat T}_{R\rightarrow L}$) is the spin-dependent transfer matrix accounting for both elastic and inelastic tunneling. In the spinor formalism, the charge current and spin current are expressed $j_e=Tr[{\hat J}]$ and ${\cal J}_s=Tr[{\bm{\hat\sigma}}{\hat J}]$, where ${\bm{\hat \sigma}}$ is the vector of Pauli spin matrices. The hat $\hat{}$ denotes 2$\times$2 matrices in spin space. This form accounts for contributions of rightward and leftward electrons originating from the left (L) or right (R) reservoirs. Using the above definition assumes that tunneling is only due to interfacial densities of states and transmission through the barrier. \par

The important point here is the definition of the transmission matrices ${\hat T}_{L\rightarrow R}$ and ${\hat T}_{R\rightarrow L}$. In the case of spin torque, we have to extend the BTM formalism to non-collinear magnetization directions. That is to say, the interfacial densities of states must be rewritten in the absolute quantization axis that we chose aligned on the magnetization of the right layer. Slonczewski \cite{slonc07} and Levy and Fert\cite{magnon} used this approach to derive explicit low bias expressions of the in-plane torque. They assumed that the ballistic tunneling through the barrier is spin independent. In other words, the spin state does not rotate during the tunneling. The spin impinging on the free layer is then simply aligned on the polarizer orientation ${\bm p}$, as depicted in Fig. \ref{fig:fig1}(b). This picture, that we refer to as the {\em rigid spin} assumption, allows to get an expression for the in-plane torque, but does not provide any indications about the nature of the out-of-plane torque. However, numerical calculations \cite{xiao,wil,manchon,theo,heiliger} and experimental evidence \cite{petit,deac,li,sankey,oh} have demonstrated the existence of a large out-of-plane component of the spin torque. This indicates that a fundamental ingredient is missing in the original BTM approach and corrections to the rigid spin assumption must be considered. Actually, when a spin-polarized electron impinges on a ferromagnet, its transmission through the interface is accompanied by a {\em spin rotation}. This can be easily understood using a toy model. Let's consider a rightward electron in the rotating frame of the right electrode. Its original wave function $|\Psi\rangle_0$ is transformed into $|\Psi\rangle_t$ under transmission through the interface
\begin{eqnarray}
|\Psi\rangle_0=\cos\frac{\theta}{2}|\uparrow\rangle-\sin\frac{\theta}{2}|\downarrow\rangle\Rightarrow |\Psi\rangle_t={\hat T}|\Psi\rangle_0=(T^\uparrow\cos\frac{\theta}{2}|\uparrow\rangle-T^\downarrow\sin\frac{\theta}{2}|\downarrow\rangle).
\end{eqnarray}
Since the interface between the barrier and the free magnetic layer is spin-dependent, it filters the electron spin and its majority and minority projections are not transmitted the same way: the interaction with the interface induces a phase shift between the majority and minority projections of the spin. In other words, since the impinging spin feels a magnetic interaction at the interface, it {\em rotates} upon transmission/reflection. This effect has been identified by Stiles and Zangwill\cite{stiles2002} in metallic spin-valves and by Manchon et al.\cite{manchon} in magnetic tunnel junctions. These references demonstrate that the supplementary angle gained over transmission can be quite large and depends on the direction of incidence. In metallic systems, averaging over the Fermi surface quenches this contribution\cite{stiles2002}, whereas in magnetic tunnel junctions, the wave vector filtering yields an effective angle that is quite large and has a component out of the (${\bm m},{\bm p}$) plane\cite{manchon} [see Fig. \ref{fig:fig1}(b)]. Consequently, the effective spin density in the free layer is rotated out of this plane and produces a so-called out-of-plane torque.\par

This implies that the tunneling matrix is no more spin independent and that the incident spin "sees" a free layer magnetization with a virtual orientation ($\theta$,$\varphi$), $\varphi$ being the phase induced by the spin-dependent transmission. The expression for the tunnel current spinor in the quantization axis of the right (free) layer is then
\begin{eqnarray}\label{eq:5}
\hat{J}=2\pi\frac{e}{\hbar}\int d\epsilon[{\cal R}\hat{\rho}_{L}\hat{T}_{L\rightarrow R}{\cal R}_{\varphi_L}\hat{\rho}_{R}{\cal R}_{\varphi_L}^\dagger\hat{T}_{L\rightarrow R}^\dagger  {\cal R}^\dagger f_L(1-f_R)-\hat{\rho}_{R}\hat{T}_{R\rightarrow L}{\cal R}_{\varphi_R}\hat{\rho}_{L}{\cal R}_{\varphi_R}^\dagger\hat{T}_{R\rightarrow L}^\dagger f_R(1-f_L)],
\end{eqnarray}
where we defined the interfacial density of state of the $i$th electrode and the rotations matrices ${\cal R}$ and ${\cal R}_\varphi$
\begin{eqnarray}
{\hat \rho}_i&=&\frac{1}{2}\left(
\begin{tabular}{cc}
$\rho_i+\Delta \rho_i$&0\\
0 & $\rho_i-\Delta \rho_i$
\end{tabular}\right),\;{\cal R}=\left(
\begin{tabular}{cc}
$\cos\frac{\theta}{2}$&$-\sin\frac{\theta}{2}$\\
$\sin\frac{\theta}{2}$ & $\cos\frac{\theta}{2}$
\end{tabular}\right),\;{\cal R}_\varphi=\left(
\begin{tabular}{cc}
$\cos\frac{\theta}{2}$&$-e^{i\varphi}\sin\frac{\theta}{2}$\\
$e^{-i\varphi}\sin\frac{\theta}{2}$ & $\cos\frac{\theta}{2}$
\end{tabular}\right).
\end{eqnarray}
As mentioned above, note that the matrices $R_{\varphi_i}$ account for the rotation of the electron spin when tunneling through the barrier whereas ${\cal R}$ ensures that the spinor current is defined in the quantization frame of the right layer. A straightforward calculation leads to
\begin{eqnarray}
{\bf T}_{||}(\epsilon)&=&2\frac{\pi}{\hbar}\int d\epsilon|T|^2\rho_R\Delta\rho_L(\cos\varphi_Lf_L-\cos\varphi_Rf_R),\\
{\bf T}_{\bot}(\epsilon)&=&2\frac{\pi}{\hbar}\int d\epsilon|T|^2(\rho_R\Delta\rho_L\sin\varphi_Lf_L+\rho_L\Delta\rho_R\sin\varphi_Rf_R),
\end{eqnarray}
At this stage we have made no assumption about the symmetry of the junction. Note that in general the correction from the interfacial precession is small $\varphi<<1$. In this case, it clearly appears that the in-plane torque is {\em antisymmetric} (in the limit of small $\varphi$) when exchanging L and R indices, whereas the out-of-plane torque remains {\em symmetric}. We considered ballistic tunneling in a thick barrier, but it is in principle possible to have asymmetric electrodes or asymmetric barrier profile. Let's now assume a symmetric tunnel junction submitted to a bias voltage. In this case, we can rewrite the explicit energy dependence of the different terms involved in the spin transport
\begin{eqnarray}\rho_i(\epsilon)=\rho(\epsilon\pm eV/2),\;\Delta\rho_i(\epsilon)=\Delta\rho(\epsilon\pm eV/2),\;\varphi_i(\epsilon)=\varphi_i(\epsilon\pm eV/2).
\end{eqnarray}
where the sign $\pm$ is associated with the left (right) electrode respectively, and the Fermi-Dirac distribution is given by $f_{L,R}(\epsilon)=(e^{(\epsilon-\epsilon_F\pm eV/2)/k_BT}+1)^{-1}$. Then, the in-plane and out-of-plane torques become\begin{eqnarray}
T_{||}&=&\int d\epsilon[ {\hat {\bf \tau}}^{||}(\epsilon+eV/2)-{\hat {\bf \tau}}^{||}(\epsilon-eV/2)],\\
T_{\bot}&=&\int d\epsilon[ {\hat {\bf \tau}}^{\bot}(\epsilon+eV/2)+{\hat {\bf \tau}}^{\bot}(\epsilon-eV/2)].
\end{eqnarray}
Therefore, independently on the detail of the band structure and in the limit of the model, if the junction is symmetric, it implies that
\begin{eqnarray}
T_{||}&=&a_1V,\label{eq:eqv}\\
T_{\bot}&=&b_0+b_2V^2,
\end{eqnarray}
in qualitative agreement with the numerical estimates \cite{xiao,wil,manchon,theo,heiliger}. Note that the actual energy dependence of the interfacial spin-dependent densities of states influences the bias dependence of the spin torque at large voltages. First, as experimentally demonstrated by Valenzuela \cite{valenzuela}, the polarization of the impinging electrons depends on the bias polarity which induces quadratic corrections to the simple linear voltage dependence of the in-plane torque presented in Eq. \ref{eq:eqv}. Second, at large bias dependence, Fowler-Nordheim processes start dominating the transport which induces oscillatory voltage dependences of the tunneling magnetoresistance and spin torque, as calculated by Tang et al. \cite{tang,montaigne}.

\subsection{Materials Consideration}

\begin{figure}[h]
\centering
\includegraphics[width=6cm]{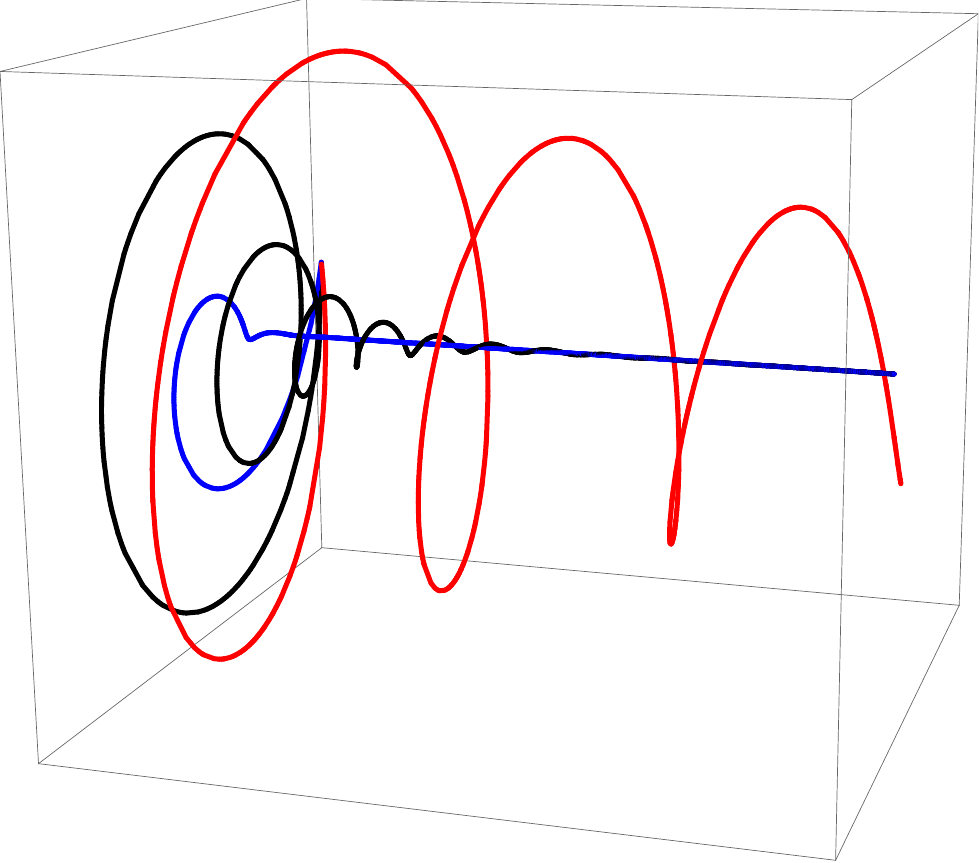}
\caption{\label{fig:fig2}(Color online) Schematics of the spatial profile of the spin density transverse to the local magnetization, extending in the free layer. We represent typical profiles in the cases of a weak ferromagnet (long precession length - red), strong ferromagnet (short precession length - black) and half-metallic free layer (exponential decay at the interface - blue).}
\end{figure}

When a spin-polarized electron enters into the free layer, its spin precesses around the local magnetization with a spatial period of $2\pi/(k^\uparrow-k^\downarrow)$ \cite{stiles2002,manchon}. Since the angle between the electron spin and the local magnetization depends on the incident angle of the electron, the precession period also depends on the angle of incidence. After averaging over the Fermi surface, as mentioned above, the effective magnetization displays a oscillatory damped spatial profile as depicted schematically in Fig. \ref{fig:fig2}.\par

The most common system investigated to date is the MgO-based magnetic tunnel junctions with CoFeB electrodes. For thick enough barriers, the electron wave function is filtered so that the transport is dominated by $\Delta_1$ symmetries \cite{butler}. Since $\Delta_1$ bands are half metallic in CoFeB, the junctions display huge tunneling magnetoresistance ratios \cite{mgo}. In this limit, the majority projection of the electron spin propagates regularly in the free layer whereas the minority spin wave function is evanescent. As a consequence, the spin density profile is heavily damped, as shown in Fig. \ref{fig:fig2}, and the spin torque is very strongly localized at the interface. Thus, the model of interfacial spin torque proposed above is valid. Note however that in thin MgO barriers, resonant states arise \cite{bonnel} that alter the tunnel magnetoresistance by allowing for minority electrons from other band symmetries to tunnel through the barrier. In this case, the minority waves are no more evanescent and are allowed to propagate in the free layer, yielding a torque that can extend through the volume of this layer.\par

\subsection{Structural Asymmetries}
\begin{figure}
\centering\begin{tabular}{cl}
\includegraphics[width=8cm]{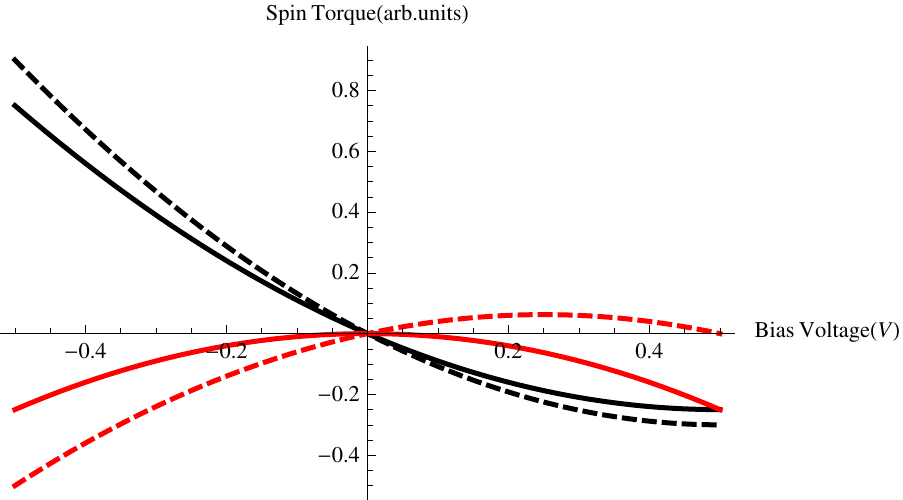}&\includegraphics[width=8cm]{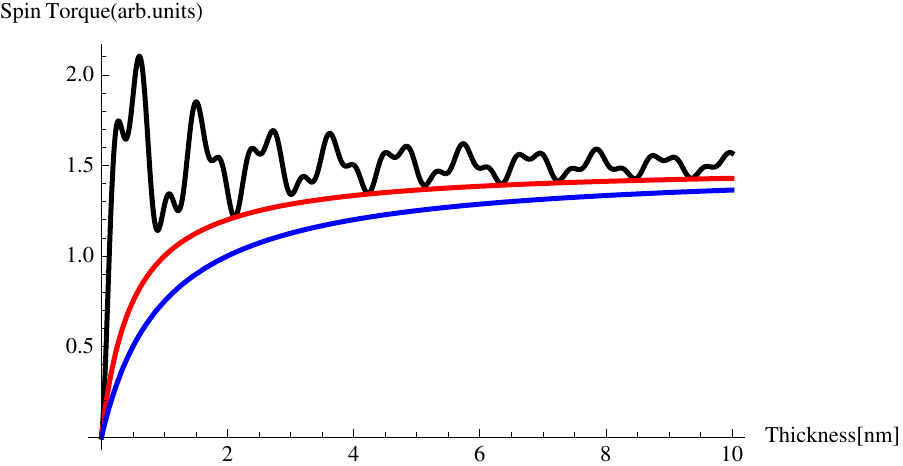}
\end{tabular}
\caption{\label{fig:fig3}(Color online) (a) Bias dependence of the in-plane (black) and out-of-plane (red) torques in the case of symmetric (solid) and asymmetric barrier (dashed). See Refs. *** for details; (b) Spin torque magnitude as a function of the free layer thickness in the case of coherent ballistic transport (quantum interferences are visible - black), spin diffusion in the short spin dephasing limit (the thickness dependence in $\propto 1/d$ - red) and long spin dephasing (deviation from $1/d$ - blue).}
\end{figure}

The previous simple model provides a qualitative description of the bias dependence expected in the case of asymmetric junctions. When the work function of the left and right electrodes are different or when the materials of the electrodes are different from each other, they induce asymmetries in the transport: the spin torque is more efficient at one bias polarity than at the other. In other words, when inserting structural asymmetries, ${\hat {\bf \tau}}_{L\rightarrow R}\neq{\hat {\bf \tau}}_{R\rightarrow L}$. The first implication is that the bias dependence of the out-of-plane torque deviates from the "conventional" quadratic dependence. This is schematically represented in Fig. \ref{fig:fig3}, left panel. In this figure, we represented the bias dependence obtained using a free electron model in the case of symmetric (solid lines) and asymmetric tunnel barrier (dashed lines). Note the presence of a quadratic component $V^2$ in the in-plane torque, that arises from the energy dependence of the interfacial densities of states. Details can be found elsewhere \cite{wil,manchon,tang}.\par

This feature has been recently exploited by Oh et al. \cite{oh} to reduce the back-hopping process observed at large biases \cite{sun}. As discussed above, in a symmetric junction, whereas the sign of the in-plane torque depends on the polarity of current injection ($\propto V$), the out-of-plane torque is always in the same direction ($\propto V^2$). Consequently, at positive polarity both torques favor the antiparallel configuration, whereas at negative polarity, the in-plane torque favors the parallel configuration while the out-of-plane torque favors the antiparallel one. This competition leads to back-hopping of the magnetization state which is detrimental for applications such as MRAMs\cite{sun,oh}. Introducing artificial asymmetries in the system, by using either different electrode materials or by engineering the barrier potential profile, Oh et al.\cite{oh} showed that it is possible to quench this back-hopping process by adding a linear part to the bias dependence of the out-of-plane torque.\par

\subsection{Interfacial Scattering}

We now briefly review previous works on the influence of interfacial spin scattering in magnetic tunnel junctions \cite{magnon,prb}. It is well known that electron interactions with thermal excitations can affect magnetic tunnel junctions properties such as the magnetoresistance \cite{zhang97}. Hot electrons-induced magnon emission and absorption affect the polarization of the tunneling electrons and electron-phonon interactions result in thermally-assisted tunneling. In the presence of electron-magnon and electron-phonon interactions, the transfer matrix becomes both spin and energy dependent
\begin{eqnarray}\label{eq:sd}
\hat{T}_{L\rightarrow R}^{e-m}&=&\hat{T}_{{L\rightarrow R}}\left(\hat{I}+\sqrt{\frac{Q^m}{N}}(\bm{{\hat\sigma}}.{\bf S}_{tr}^R+\bm{{\hat\sigma}}.{\bf S}_{tr}^L)\right),\\\label{eq:ph}
\hat{T}_{L\rightarrow R}^{e-ph}&=&\hat{T}_{{L\rightarrow R}}\left(1+\sqrt{\frac{Q^{ph}_{\bf q}}{N}}(b_{\bf q}+b_{\bf q}^+)\right)\hat{I},
\end{eqnarray}
where $Q^m$ ($Q^{ph}_{\bf q}$) is the phenomenological electron-magnon (electron-phonon)
efficiency, $N$ is the number of atoms per cell, $\bm{\sigma}$ is
the vector of Pauli spin matrices and ${\bf S}_{tr}^{L(R)}$ are the
transverse part of the magnetizations of the left and right
electrodes. Details about the derivation of Eqs. (\ref{eq:sd})-(\ref{eq:ph}) can be found in Ref. [25]. Assuming that the electron spin-dependent densities of state do not vary much over the range $eV$, and considering acoustic phonons ($\omega \propto q$, $Q_{\bf q}\propto q$) with a density of states of the form $\rho_{ph}(\omega)\propto \omega^{\nu}$, we obtain, at T=0 K and low bias voltage
\begin{eqnarray}
&&T_{||}=G_0P^L\sin\theta(1+\zeta_{ph}|V|^{\nu+2})V,\label{eq:17}\\
&&T_\bot-T_{\bot0}=G_0P^R\phi_L\sin\theta\zeta_{ph}|V|^{\nu+3},\label{eq:18}
\end{eqnarray}
$\zeta_{ph}$ being a coefficient that depends on the electron-phonon coupling, Fermi energy, Debye temperature $\Theta_D$ etc... Note that the symmetry of the out-of-plane torque against the bias is conserved, whereas the in-plane torque acquires an antisymmetric component. \par

In the case of electron-magnon interaction, the transfer matrix [Eq. (\ref{eq:sd})] possesses non-diagonal elements that are responsible for spin-flip scattering. We then expect a much more complex influence on the torque. Assuming a magnon density of states of the form $\rho_m(\omega)=\omega^\nu$, symmetric electrodes ($P^i=P$, $\varphi_i=\varphi_R$) and T=0 K, we find:
\begin{eqnarray}
&&T_{||}-T_{||0}\propto \sin\theta[P(1+P)-(1-P)(1+P\cos\theta)] V^{\nu+2},\label{eq:19}\\
&&T_{\bot}-T_{\bot0}\propto P\phi\sin\theta(1-\cos\theta) |V|^{\nu+2}.\label{eq:20}
\end{eqnarray}
The detail of these expressions can be found in Ref. [25]. Interestingly the out-of-plane torque and the conductance (not shown) both acquire a component that is symmetric against the bias because neither electron-magnon nor electron-phonon scattering break the junction symmetry. Furthermore, since the electron-magnon interaction mixes the majority and minority channels, the angular dependence is also affected, contrary to the case of electron-phonon coupling.

\subsection{Finite thicknesses in ballistic regime}

In the case where the electrodes' thickness is much larger than the spin dephasing length, the incident spin current is totally absorbed in the free layer. Therefore, the torque is determined by the interfacial spin current only. For example, in the case of half-metallic transport of $\Delta_1$ electrons in FeCo/MgO/FeCo junctions, since $\Delta_1$ minority electrons cannot propagate into the free layer, the length over which spin precesses (or in other words, the length over which spin transfer takes place) is reduced to one to two monolayers \cite{heiliger}. In this case, the total torque exerted on the layer does not depend on the layer thickness since all the injected transverse spin current has been absorbed at the interface. Therefore, the average torque acting on the free layer is inversely proportional to the layer thickness ($\propto 1/d$). \par

In the case of normal metallic behavior, both majority and minority spin propagate in the free layer, giving rise to a non-vanishing coherence length (see Fig. 2). In the case of free layers of thicknesses comparable to the spin precession length, the transverse spin current is not fully absorbed in the free layer and is {\em reflected} by the second interface. This induces  quantum interferences between the injected spin current and the reflected (counter propagating) one. The precession pattern of the spin density is therefore a combination between $\exp[i(k^\uparrow-k^\downarrow)x]$ and  $\exp[i(k^\uparrow+k^\downarrow)x]$. This interference pattern is therefore affected by the thickness of the free layer and produces a complex thickness dependence, as schematically depicted in Fig. \ref{fig:fig3}(b). This dependence has been studied numerically in several references \cite{wil,heiliger}. Note that quantum coherent is seminal to obtain this pattern and interfacial roughness or impurity scattering can easily destroy these interferences.\par

\section{DIFFUSION OF TUNNELING SPIN TORQUE}

In the present section, we briefly present recent results obtained in the opposite limit of diffusive transport in the free layer \cite{manchon2012}. As seen in the previous section, in the coherent ballistic regime, quantum interferences occur when the free layer thickness is finite. However, in the limit of diffusive transport, quantum interferences are quenched and spin relaxation and diffusion dominate the transport. The tunneling process through the insulating barrier imposes a ballistic injection of carriers at the interface between the tunnel barrier and the free layer which constitutes a {\em boundary condition} to the coupled diffusive spin transport equations for the transverse component of the spin accumulation vector $\bm{s}$. Along these guidelines, the spin dynamics of the transverse spin accumulation in the free layer is governed by the following coupled equations in steady state
\begin{eqnarray}
{\bm \nabla}\cdot{\cal J}&=&-\frac{1}{\tau_J}{\bm s}\times{\bm m}-\frac{1}{\tau_\phi}{\bm m}\times({\bm s}\times{\bm m})-\frac{1}{\tau_{sf}}{\bm s},\label{eq:s1}\\
{\cal J}&=&-{\cal D}{\bm\nabla}\otimes{\bm s},\label{eq:s2}
\end{eqnarray}
where ${\bm s}$ is the spin accumulation, ${\bm m}$ is the direction of the local magnetization, and ${\cal J}$ is the spin current tensor. The diffusion is characterized by the diffusion constant $\cal{D}$ and the spin dynamics is controlled by the spin precession time $\tau_J$, spin decoherence time $\tau_{\phi}$ and spin relaxation time $\tau_{sf}$. Whereas the spin relaxation affects the three spin components, the spin precession and spin decoherence terms only affect the two transverse components of the spin accumulation vector. The spin torque is defined as the spatial change of spin current, compensated by the spin relaxation term
\begin{eqnarray}
&&{\bf T}=\frac{1}{\Omega}\int_{\Omega}d\Omega\left(-{\bm \nabla}\cdot{\cal J}-\frac{1}{\tau_{sf}}{\bm s}\right).
\end{eqnarray}
Here, $\Omega$ is the volume of the magnetic layer. Note the seminal difference with the definition exploited in the previous section: the spatial variation of the spin current is now partially balanced by the loss of the spin angular momentum through spin flip scattering and the electron is now propagating {\em diffusively} in the free layer yielding a specific spatial distribution of the spin current that needs to be evaluated.\par

\begin{figure}[h!]
\centering
\includegraphics[width=12cm]{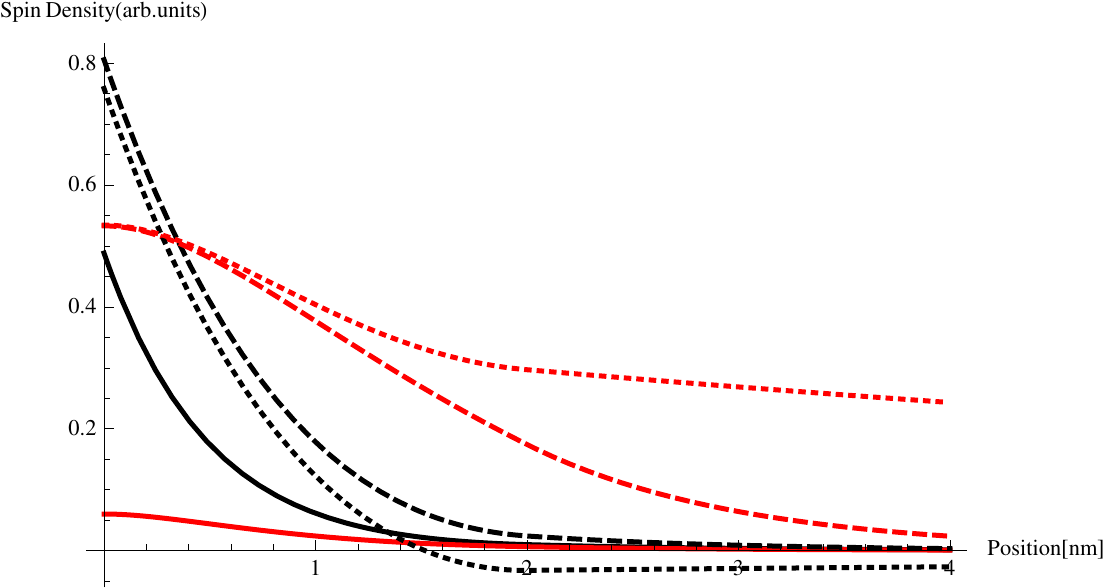}
\caption{\label{fig:fig4}(Color online) Spatial distribution of the out-of-plane (black) and in-plane (red) components of the spin accumulation in the free layer/capping layer bilayer structure for short spin dephasing length ($\lambda_\phi=0.5$nm - solid lines), long spin dephasing length ($\lambda_\phi=1.5$nm - dashed lines) for short spin relaxation length in the capping layer ($\lambda_{sf}^N=1$nm). The dotted lines display the spin accumulation profile for long spin relaxation length in the capping layer  ($\lambda_\phi=1.5$nm and $\lambda_{sf}^N=10$nm). The interface with the tunnel barrier is located at $x=0$ and the interface between the free layer and the capping layer is at $x=2$nm. These curves were calculating following the theory developed in Ref. 34.}
\end{figure}

We consider a finite free layer embedded between the tunnel barrier and a normal metallic capping layer. Assuming that a spin current ${\cal J}_0={\cal J}_{\|}+i{\cal J}_\bot$ is imposed by the ballistic tunnel at the interface with the barrier, and considering a standard continuity condition at the interface with the capping layer \cite{manchon2012}, the total spin torque exerted on the ferromagnet is
\begin{eqnarray}
T_\|+iT_\bot=\frac{{\cal J}_{0}}{d}\frac{L^2}{L_0^2}\frac{\cosh\frac{d}{L}+\eta\sinh\frac{d}{L}-1}{\cosh\frac{d}{L}+\eta\sinh\frac{d}{L}},\label{eq:12}\end{eqnarray}
where $\frac{1}{L_0^2}=-\frac{i}{\lambda_J^{2}}+\frac{1}{\lambda_\phi^2}$, $\frac{1}{L^2}=-\frac{i}{\lambda_J^{2}}+\frac{1}{\lambda_\phi^2}+\frac{1}{\lambda_{sf}^2}$, $d$ is the free layer thickness and $\eta$ is a parameter that only depends on the bulk characteristics of the free and capping layers. We clearly see that in general the in-plane and out-of-plane torques are a mixture between in-plane and out-of-plane interfacial spin currents
\begin{eqnarray}
&&T_\|=\alpha {\cal J}_\|+\beta{\cal J}_\bot,\\
&&T_\bot=-\beta {\cal J}_\|+\alpha{\cal J}_\bot,\label{eq:13}\end{eqnarray}
These expressions reduce to the ballistic case studied in the previous section in the case of infinitely free layer thickness ($d\rightarrow\infty$) and infinite spin diffusion length ($\lambda_{sf}\rightarrow\infty$). However, in the most general case, it appears that the spin torque can not be simply identified to the interfacial spin current. This property arises from the fact that (i) spin relaxation is always present, alters the propagation of the spin density in the metallic layers and {\em redistributes} the spin degree of freedom on the two torque components; (ii) when the free layer thickness is comparable to the spin dephasing length, the transverse spin density diffuses towards the interface with the capping layer which in turns influences the spin dynamics in the free layer. This last effect virtually enhances the impact of the spin diffusion length on the spin current mixing. These two situations are illustrated in Fig. \ref{fig:fig4}.\par

The diffusion of spin current in the free layer has two major implications: the {\em redistribution} of the spin angular momentum is thickness-dependent and as soon as the thickness of the free layer exceeds the spin dephasing length, the interfacial spin current limit is recovered [see Fig. \ref{fig:fig3}(b), red line]. Note that the deviation we obtain here is not related to quantum interferences (which are destroyed by impurity scattering and roughness), but rather to the incomplete absorption of the spin current: when the thickness of the free layer is smaller than the spin dephasing length $\lambda_\phi$, the transverse spin current responsible for the spin torque is not fully absorbed in the free layer and diffuses towards the capping layer. This induces a deviation from the usual 1/d-thickness dependence [see Fig. \ref{fig:fig3}(b), blue line]. Increasing the thickness of the free layer improves the absorption of the spin current and for thicknesses much larger than the spin dephasing length, the thickness dependence of the torque recovers the 1/d limit. Note that in the case of half-metallic behavior, as in Fe/MgO/Fe tunnel junctions, the minority band of $\Delta_1$ symmetry does not propagate in the ferromagnet which induces a quenching of the spin dephasing length and reducing the spin torque to a 1/d behavior.\par

Second, this new dynamics mixes the two transverse components of the spin current and is therefore responsible for the deviation from the ballistic bias dependence shown in Eq. (\ref{eq:2}). If one assumes a bias dependence of the spin current on the form ${\cal J}_{\|}=a_1V$ and ${\cal J}_{\bot}=b_2V^2$, as expected and observed for systems such as Fe/MgO/Fe \cite{sankey,heiliger}, then both in-plane and out-of plane spin torque components will be a mixture of linear and quadratic bias dependences
\begin{eqnarray}
&&T_\|=\alpha a_1V+\beta a_2V^2,\\
&&T_\bot=-\beta a_1V+\alpha a_2V^2,\label{eq:13}\end{eqnarray}

This spin mixing may explain the recently obtained experimental results that show a strong linear voltage dependence of the out-of-plane torque, in contrast with the predictions \cite{theo,wil,xiao,manchon}.
\section{CONCLUSION}
In conclusion, the role of spin diffusion in the metallic layers of MTJs has been addressed theoretically. Assuming an interfacial bias-driven spin current at the interface between the insulator and the ferromagnet, the spin diffusion equation is solved and describes a complex spin dynamics in the metallic layers. It is found that this dynamics mixes the components of the spin current tranverse to the local magnetization which results in a superposition between linear and quadratic bias dependence of the out-of-plane torque. The thickness dependence of the spin transfer torque is also altered for small thicknesses.\par
\acknowledgments     

The author acknowledge S. Zhang, K.-J. Lee, J. Grollier, H. Jaffr\`es, R. Matsumoto and A. Fert for stimulating discussions and fruitful collaborations.



\begin{thebibliography}{1}
\bibitem{reviewchappert} C. Chappert, A. Fert and F. Nguyen Van Dau, Nature Materials {\bf6}, 813 (2007).
\bibitem{ieee} S. Ikeda, J. Hayakawa, Y. M. Lee, F. Matsukura, Y. Ohno, T. Hanyu, and H. Ohno, IEEE Trans. Elec. Dev. {\bf54}, 991 (2007).
\bibitem{tas} I.L. Prejbeanu, M. Kerekes, R. C. Sousa, H. Sibuet, O. Redon, B. Dieny, and J. P. Nozieres, J. Phys. Condensed Matter 19, 165 (2007).
\bibitem{Slonc96} J. C. Slonczewski, J. Magn. Magn. Mater. {\bf159}, L1 (1996); L. Berger, Phys. Rev. B {\bf54} 9353, (1996).
\bibitem{review1} D. C. Ralph and M. D. Stiles, J. Magn. Magn. Mater. {\bf 320}, 1190-1216 (2008).
\bibitem{reviewmtj} J.Z. Sun and D.C. Ralph, J. Magn. Magn. Mater. {\bf320}, 1227 (2008).
\bibitem{chapter} A. Manchon, and S. Zhang, 'Spin Torque in Magnetic Systems: Theory', Handbook of Spin Transport and
Magnetism, Eds. E.-Y. Tsymbal and I. Zutic, Chap. 8, CRC Press, August 2011.
\bibitem{huai} J.Z. Sun, J. Magn. Magn. Mater. {\bf202}, 157 (1999); Y. Huai \etal, Appl. Phys. Lett. {\bf84}, 3118
(2004); G. D. Fuchs \etal, Appl. Phys. Lett. {\bf85}, 1205 (2004); D. Chiba \etal, Phys. Rev. Lett. {\bf93}, 216602 (2004).
\bibitem{sankey} J. C. Sankey \etal, Nature Physics {\bf4}, 67
(2008); H. Kubota \etal, Nature Physics {\bf4}, 37 (2008).
\bibitem{theo} I. Theodonis \etal, Phys.Rev. Lett. {\bf97}, 237205 (2006).
\bibitem{slonc07} J. C. Slonczewski, Phys. Rev. B {\bf 71}, 024411 (2005); J.C. Slonczewski and J.Z. Sun, J. Magn. Magn. Mater.
{\bf310}, 169-175 (2007); See also, J. C. Slonczewski, Phys. Rev. B {\bf 39}, 6995 (1989).
\bibitem{manchon} A. Manchon \etal, J. Phys.: Condens. Matter {\bf20}, 145208 (2008); {\em ibid} {\bf19}, 165212 (2007).
\bibitem{xiao} J. Xiao, G. E. W. Bauer, and A. Brataas, Phys. Rev. B {\bf77}, 224419 (2008).
\bibitem{wil} M. Wilczynski, J. Barnas, and R. Swirkowicz, Phys. Rev. B {\bf77}, 054434 (2008).
\bibitem{heiliger} C. Heiliger and M. D. Stiles, Phys. Rev. Lett. {\bf100}, 186805 (2008); X. Jia, K. Xia, Y. Ke, and H. Guo, Phys. Rev. B {\bf84}, 014401 (2011).
\bibitem{tang} Y.-H. Tang \etal, Phys. Rev. Lett. {\bf103}, 057206 (2009); Phys. Rev. B {\bf81}, 054437 (2010).
\bibitem{oh} S.-C. Oh \etal, Nature Physics {\bf 5}, 898 (2009).
\bibitem{deac} A. M. Deac \etal, Nature Physics {\bf4}, 803 (2008).
\bibitem{petit}S. Petit \etal, Phys. Rev. Lett. {\bf98}, 077203 (2007); O. G. Heinonen, S. W. Stokes, and J. Y. Yi, Phys. Rev. Lett. {\bf105}, 066602 (2010).
\bibitem{li} Z. Li \etal, Phys. Rev. Lett. {\bf100}, 246602 (2008).
\bibitem{prb} A. Manchon, S. Zhang and K.-J. Lee, Phys. Rev. B {\bf82}, 174420 (2010).
\bibitem{bardeen} J. Bardeen, Phys. Rev. Lett. {\bf6}, 57 (1961). 
\bibitem{julliere} M. Julliere, Phys. Lett. A {\bf54}, 225 (1975).
\bibitem{nozieres} C. Caroli, R. Combescot, P. Nozieres, and D. Saint-James, J. Phys. C {\bf4}, 916 (1971).
\bibitem{magnon} P. M. Levy and A. Fert, Phys. Rev. Lett. {\bf97}, 097205 (2006); Phys.Rev. B {\bf74}, 224446 (2006); A. Manchon and S. Zhang, Phys. Rev. B {\bf79}, 174401 (2009).
\bibitem{stiles2002} M. D. Stiles and A. Zangwill, Phys. Rev. B {\bf66}, 014407 (2002).
\bibitem{valenzuela} S. O. Valenzuela, D. J. Monsma, C. M. Marcus, V. Narayanamurti, and M. Tinkham, Phys. Rev. Lett. {\bf94}, 196601 (2005).
\bibitem{montaigne} See also F. Montaigne, M. Hehn, and A. Schuhl, Phys. Rev. B {\bf64}, 144402 (2001).
\bibitem{butler} W. H. Butler, X.-G. Zhang, T. C. Schulthess, J. M. MacLaren, Phys. Rev. B {\bf63}, 054416 (2001).
\bibitem{mgo} S.S.P. Parkin, et al. Nature Mater. {\bf3}, 862Ð867 (2004); S. Yuasa, T. Nagahama, A. Fukushima, Y. Suzuki, and K. Ando, Nature Mater. {\bf3}, 868Ð871 (2004).
\bibitem{bonnel} J. M. Teixeira, J. Ventura, J. P. Araujo, J. B. Sousa, P. Wisniowski, S. Cardoso and P. P. Freitas, Phys. Rev. Lett. {\bf106}, 196601 (2011); F. Bonell, T. Hauet, S. Andrieu, F. Bertran, P. Le F\`evre, L. Calmels, A. Tejeda, F. Montaigne, B. Warot-Fonrose, B. Belhadji, A. Nicolaou, and A. Taleb-Ibrahimi, Phys. Rev. Lett. {\bf108}, 176602 (2012).
\bibitem{sun}J. Z. Sun \etal, J. Appl. Phys. {\bf105}, 07D109 (2009); T. Min \etal, J. Appl. Phys. {\bf105}, 07D126 (2009).
\bibitem{zhang97} S. Zhang, P. M. Levy, A. C. Marley, and S. S. P. Parkin, Phys. Rev. Lett. {\bf79}, 3744 (1997); J. S. Moodera, J. Nowak, and R. J. M. van de Veerdonk, Phys. Rev. Lett. {\bf80}, 2941 (1998).
\bibitem{manchon2012} A. Manchon, R. Matsumoto, H. Jaffr\`es and J. Grollier, arxiv:1204.5000 (2012).
\end{thebibliography}
\end{document}